\documentclass[12pt]{emulateapj}
\usepackage[]{natbib}
\usepackage{graphicx, amsmath}
\def\etal{{et~al.\null}}
\def\eg{{e.g.,\ }}

\begin{document}

\title{The Dust Attenuation Curve versus Stellar Mass for Emission Line Galaxies at $z \sim 2$} 
\shorttitle{The $z \sim 2$ Dust Attenuation Relation}
\author{Gregory R. Zeimann\altaffilmark{1,2},
Robin Ciardullo\altaffilmark{1,2},
Caryl Gronwall\altaffilmark{1,2},
Joanna Bridge\altaffilmark{1,2},
Hunter Brooks\altaffilmark{1},
Derek Fox\altaffilmark{1,2},
Eric Gawiser\altaffilmark{3},
Henry Gebhardt\altaffilmark{1,2},
Alex Hagen\altaffilmark{1,2},
Donald P. Schneider\altaffilmark{1,2},
Jonathan R. Trump\altaffilmark{1,2,4}}
\altaffiltext{1}{Department of Astronomy \& Astrophysics, The Pennsylvania
State University, University Park, PA 16802}
\altaffiltext{2}{Institute for Gravitation and the Cosmos, The Pennsyslvania State University, University Park, PA 16802}
\altaffiltext{3}{Department of Physics and Astronomy, Rutgers, The State University of New Jersey, Piscataway,NJ 08854}
\altaffiltext{4}{Hubble Fellow}
\email{grzeimann@psu.edu}

\begin{abstract}

We derive the mean wavelength dependence of stellar attenuation in a sample of 239 high
redshift ($1.90 < z < 2.35$) galaxies selected via {\it Hubble Space Telescope} 
({\it HST}) WFC3 IR grism observations of their rest-frame optical emission lines.  
Our analysis indicates that the average reddening law follows a form similar to that derived by Calzetti et al. for local starburst galaxies.  
However, over the mass range $7.2 \lesssim \log M/M_{\odot} \lesssim 10.2$, the slope
of the attenuation law in the UV is shallower than that seen locally, and
the UV slope steepens as the mass increases. These trends are in qualitative agreement with Kriek \& Conroy, 
who found that the wavelength dependence of attenuation varies
with galaxy spectral type.  However, we find no evidence of an extinction ``bump'' at 2175~\AA\ 
in any of the three stellar mass bins, or in the sample as a whole.
We quantify the relation between the attenuation curve
and stellar mass and discuss its implications.
\end{abstract}


\section{Introduction}
\label{sec:Intro}

One of the most important uncertainties associated with the interpretation of 
high-redshift galaxy observations is our lack of understanding about interstellar 
extinction.  Measurements of galactic star formation rates (SFRs), mass-specific star 
formation rates (sSFR), age, metallicity, and size all depend in some way on the 
assumed wavelength dependence and normalization of the dust attenuation law, yet our 
knowledge of reddening and extinction remains quite limited.  From observations of 
the local universe, it is clear that reddening curves are not universal: while most 
sight-lines in the Milky Way (MW) have a strong extinction bump near 2175~\AA\ 
\citep{stecher65, cardelli+89}, the feature is weaker in the Large Magellanic Cloud 
\citep[LMC;][]{koornneef+81} and largely absent in the Small Magellanic Cloud 
\cite[SMC;][]{prevot+84}. This diversity is also reflected by systematic
changes in the far ultraviolet (UV), where the wavelength dependence of extinction
steepens dramatically as one goes from the Milky Way to the SMC\null.

Further complicating the issue is the fact that dust within a galaxy is spatially 
distributed, so simple screen models do not apply to observations of integrated 
galactic light.  As first reported by \citet{charlot+00}, patchy absorption that is 
distributed throughout a galaxy can result in an effective attenuation law that has a
much weaker wavelength dependence than that measured for stars in the Milky Way
or Magellanic Clouds.  Indeed, by observing 39 nearby starburst systems, 
\citet{calzetti+94, calzetti+00} defined an attenuation law between 1200~\AA\ and 
6300~\AA\  that is much shallower than any screen model, and that has
a notable absence of the 2175~\AA\ bump. 

This \citet{calzetti+00} law, of course, is an empirical relation which averages over much 
of the galactic physics that affects the reddening curve.   Consider that all young
stars spend the first 2 or 3 Myr of their lives enshrouded in dusty molecular clouds 
\citep{lada+03}.  The dust eventually disperses, so by the time the stars are 
$\sim 100$~Myr old, they have become dynamically mixed among the older populations.   
As a result, one often models a galaxy's dust attenuation law with two separate 
components:  an optically thin term that is applicable to the older stars, and a dense,
optically thick birth cloud that attenuates younger objects \citep[e.g.,][]
{silva+98, charlot+00}.  Using the dust models of \citet{silva+98}, \citet{granato+00} 
showed that this age dependency plays a role in defining a galaxy's net 
attenuation.   In systems dominated by young stars embedded in molecular clouds (i.e., 
starburst galaxies), a shallower UV attenuation curve is found, and no 2175~\AA\ 
bump is visible.  Conversely, in more ``normal'' star-forming galaxies, where a 
significant fraction of UV photons are emitted by stars that are no longer in their
birth clouds, the wavelength dependence is steeper, and there is evidence for excess
extinction at 2175~\AA\null.  Other studies support this model, demonstrating that 
the 2175~\AA\ bump is typically visible in ``normal'' star-forming galaxies 
\citep[e.g.,][]{burgarella+05, conroy+10}, but not in starburst systems \citep{calzetti+94}.     

In practice, age and population dependencies are rarely considered when applying
dust attenuation to distant galaxies.  Instead, the simple \citet{calzetti+00} attenuation 
law, without a 2175~\AA\ bump, is most often used to de-redden galactic
spectrophotometry in the $z \gtrsim 2$ universe \citep[e.g.,][]{bouwens+09, finkelstein+15}. 
Such an application can be justified by the fact that many of these distant systems have 
star-formation rates similar to today's most extreme starburst galaxies.   But how reliable 
is this curve?  A number of $z \gtrsim 1$ studies have presented evidence that contradicts 
the  \citet{calzetti+00} law \citep{buat+11, buat+12, scoville+15}, observing both a 
2175~\AA\ bump and a steeper wavelength dependence in the UV, and \citet{reddy+15} has
argued that, while the attenuation curve may agree with \citet{calzetti+00} at
$\lambda < 2500$~\AA, it more closely follows an SMC extinction curve at longer 
wavelengths.  Moreover, the attenuation curve at high redshift may not be universal.
\citet{kriek+13} report that, for bright ($R < 25$), composite-spectrum galaxies with
photometric redshifts between $0.5 < z < 2.0$, H$\alpha$ equivalent widths correlate 
inversely with the strength of the 2175~\AA\ bump and the slope of the UV reddening law. 
Similarly, by examining subgroups of 751 galaxies between $0.9 < z < 2.2$, \citet{buat+12} 
found that the strength of 2175~\AA\ bump was independent of SFR, but was a weak function of 
stellar mass and specific star formation rate.
 
To make further progress on this question, large samples of galaxies are needed with 
well-defined redshifts, comprehensive rest-frame UV through far-IR photometry, 
intrinsic spectral energy distributions (SEDs) that are reasonably well known, and 
stellar masses and star formation rates that span a large dynamic range.  Such samples 
now exist in the GOODS \citep{GOODS} and COSMOS \citep{COSMOS} fields, thanks 
to a combination of {\sl HST\/} infrared grism spectroscopy \citep{3DHST, AGHAST},
{\sl HST\/} high-resolution imaging, and a large database of broad- and intermediate-band 
photometry \citep{skelton+14}.  Using these data, it is possible to investigate the 
redshift dependence of dust attenuation and emission in a self-consistent manner, taking 
into account variations in the underlying stellar population and information regarding 
galactic orientation \citep{brooks+16}.  Unfortunately, such an analysis comes at a 
cost:  to compute these properties and derive realistic uncertainties, one needs to 
consider a large number of parameters.  Here, we take a more limited approach, and,
following the strategy taken by \citet{scoville+15}, 
consider only the effects of dust attenuation in a subset of $z \sim 2$ galaxies with 
intense star formation.  Specifically, by selecting galaxies with strong emission lines, 
we limit our analysis to objects whose intrinsic rest-frame UV SEDs are only weakly sensitive 
to the details of the stellar population \citep{calzetti01}.  Moreover, by including only sources
with at least two emission features, we ensure that the redshift assigned to 
each object is unambiguous.  

In this paper, we derive the mean empirical reddening curve for a sample of $1.9 < z < 2.35$ galaxies selected via the luminosity of their rest-frame optical emission lines and quantify how the properties of this curve change as a function of stellar mass.  This will lay the groundwork for a more in depth analysis which will self-consistently examine the reddening curves of individual $z \sim 2$ systems \citep{brooks+16}.
In \S\ref{sec:Sample}, we define our sample, briefly describe our modeling of the
galaxies' spectral energy distributions, and outline the procedures used to measure the
objects' stellar masses and star formation rates.  In this section,  we also show that the
UV SEDs of our objects can be well-matched using constant star-formation stellar population
models with ages between 10~Myr and 500~Myr.
In \S\ref{sec:photometry} we create our database of photometric observations by 
interpolating in the PSF-matched photometry catalog of \citet{skelton+14}, and in 
\S\ref{sec:dustlaw} we use these data to derive a mean attenuation curve for our sample 
of emission-line galaxies.  The primary result of this paper is then given in 
\S\ref{sec:dustlawmass}, where we describe how the mean attenuation curve varies with stellar 
mass, and present a prescription for correcting $z \sim 2$ emission-line galaxies for the effects 
of internal stellar reddening.   We conclude by discussing the implications of our measurements.

For this paper, we assume a $\Lambda$CDM cosmology, with $\Omega_{\Lambda} = 0.7$, 
$\Omega_M = 0.3$ and $H_0 = 70$~km~s$^{-1}$~Mpc$^{-1}$ \citep{hinshaw+13}.

\section{The Sample}
\label{sec:Sample}

To perform our analysis, we used a sample of $z \sim 2$ galaxies observed in the 
GOODS-N, GOODS-S, and COSMOS regions with the G141 near-IR grism of the 
{\sl Hubble Space Telescope's\/} Wide Field Camera 3\null.  This dataset, which was taken 
as part of the 3D-HST \citep{3DHST} and AGHAST \citep{AGHAST} surveys (GO 
programs 11600, 12177, and 12328), consists of $R \sim 130$ slitless spectroscopy over 
the wavelength range $1.08~\mu{\rm m} < \lambda < 1.68~\mu$m, and records total 
emission  line fluxes over 350~arcmin$^2$ of sky.  Tens of thousands of spectra are  
observable on these images, but of special interest to us are those produced by galaxies in the 
redshift range $1.90 < z < 2.35$, where the emission lines of [O~II] $\lambda 3727$, 
H$\beta$, and the distinctively-shaped [O~III] blended doublet $\lambda\lambda 4959,5007$
are simultaneously present in the bandpass. For these objects, unambiguous redshifts
are obtainable to an accuracy of $\Delta z \sim 0.005$ \citep{colbert+13} and total H$\beta$ 
fluxes can be measured to a 50\% completeness flux limit of 
$F \sim 10^{-17}$~ergs~cm$^{-2}$~s$^{-1}$ \citep{zeimann-1}.
Deep X-ray stacks from the surveys of \citet{elvis+09}, \citet{alexander+03} and 
\citet{xue+11} confirm that the vast majority of these objects are normal galaxies with
little evidence of nuclear activity \citep{zeimann-1}, and any source projected within 
$2\farcs 5$ of a cataloged X-ray position has been excluded from our 
analysis.  

Our parent sample is derived from the 256 galaxies analyzed by \citet{gebhardt+15}, with
a subset of 17 galaxies excluded due to their poor or inconsistent photometry ( caused mostly by 
the effects of close neighbor comtamination).  This dataset of F140W $<26$AB is well characterized, 
and has already been used for a number of studies, including the analysis of
the epoch's star formation rate indicators \citep{zeimann-1}, Ly$\alpha$
escape fraction \citep{ciardullo+14}, [Ne~III] emission \citep{zeimann-2}, and
stellar mass-metallicity-star-formation rate relation \citep{gebhardt+15}.  
To obtain the spectral energy distributions of these galaxies, we employed the 
SExtractor-based photometric catalog of \citet{skelton+14}, which combines deep,
co-added F125W + F140W + F160W images from {\sl HST\/} with the results of 30 distinct 
ground- and space-based imaging programs. The result is a homogeneous, PSF-matched set 
of broad- and intermediate-band flux densities covering the wavelength range 
$0.35~\mu$m to $8.0~\mu$m over the entire region surveyed by the {\sl HST\/} grism. In 
the COSMOS field, this dataset contains photometry in 44 separate bandpasses, with 
measurements from {\sl HST, Spitzer,} Subaru, and a host of smaller ground-based 
telescopes. In GOODS-N, the data sources are {\sl HST, Spitzer,} Keck, Subaru, and the 
Mayall telescope, and include 22 different bandpasses, while in GOODS-S, six different 
telescopes, including {\sl HST, Spitzer,} the VLT, and Subaru, provide flux densities 
in 40 bandpasses. 

\subsection{Stellar Masses and Star Formation Rates}
\label{sec:masssfr}

The stellar masses of the 239 galaxies in our sample were taken from \citet{gebhardt+15}, who used 
the Markov Chain Monte Carlo code {\tt GalMC} \citep{acquaviva+11} to model the galaxies' 
spectral energy distributions.  In brief, the \citet{skelton+14} photometry was fit to 
the 2007 version of the population synthesis models of \citet{bc03}, using a 
\citet{kroupa+01} initial mass function (IMF) over the range $0.1 \, M_{\odot} < M < 100 \,
M_{\odot}$, and a prescription for nebular continuum and emission lines given by \citet{acquaviva+11}, as 
updated by \citet{acquaviva+12}.  Since stellar abundances are poorly constrained by 
broadband SED measurements, the metallicity of our models was fixed at $Z = 0.2 \, 
Z_{\odot}$, which is close to the median gas-phase metallicity of our sample 
\citep{gebhardt+15}.  To avoid the emission from polycyclic aromatic hydrocarbons, all 
data points redward of rest-frame $3.3~\mu$m were excluded from the fits 
\citep{tielens08}.   Similarly, we removed any data with a central wavelength blueward of 1250~\AA, 
where the contribution from Ly$\alpha$ emission is uncertain, and statistical 
corrections for Lyman-line absorptions \citep{madau+95} may not always be appropriate.
Finally, for simplicity, the SFR of each galaxy was treated as
constant with time, and the attenuating effects of dust were modeled using the
prescription of \citet{calzetti+00}.  Although these latter two assumptions 
may not hold in detail, neither strongly affects our estimates for stellar masses 
\citep{conroy13, kriek+13, reddy+15, gebhardt+15}.

The values for the galactic star formation rates are given in \citet{zeimann-1}, who
applied the SFR calibrations given by \citet{kennicutt+12} to measurements of
both the objects' rest-frame continuum flux density and total monochromatic H$\beta$ emission.   
These data demonstrate that at $z \sim 2$, H$\beta$-based star formation rates 
are systematically larger than those derived from the UV continuum by a factor of $\sim 1.8$, 
and that the discrepancy cannot be due to the effects of attenuation.  \citet{zeimann-1} 
concluded that the most likely reason for the offset was metallicity, as metal-poor stars have 
lower opacities, hotter effective temperatures, and more flux beyond the ionization edge of
hydrogen (13.6~eV\null) than metal-rich stars.  
The result is a higher rate of photo-ionization in the surrounding
ISM and stronger Balmer recombination lines.  Consequently, even though UV flux
measurements integrate star formation over a longer timescale than does Balmer emission,
its use is likely to yield a more reliable estimate of
SFR than H$\beta$, which is usually weakly detected and susceptible to the effects of
underlying Balmer absorption.

Of course, to derive UV-based SFRs, \citet{zeimann-1} had to adopt some estimate
of internal extinction.  They assumed that the intrinsic rest-frame UV
continua of the galaxies is well-represented by a power law, $F_{\lambda} \propto 
\lambda^{\beta_0}$,  with $\beta_0 = -2.25$.  Any flux distribution flatter than this was
attributed to internal reddening, with $A_{1600} = 2.31 \Delta \beta$ \citep{calzetti01}.  
\citet{zeimann-1} remarked that while their best-fit relationships between the
UV and H$\beta$ extinctions were consistent with the \citet{calzetti01} law, other 
formulations with different slopes and intercepts were still possible.   For the purpose of 
understanding the effective ages of our systems only, we use the \citet{zeimann-1} 
UV SFRs in our analysis.  If we were to transform these values into H$\beta$-based SFRs or apply a different extinction law,
the conclusions of this paper would remain unaffected, as the range of ages used in our SED modeling analysis is 
quite conservative (see \S~\ref{sec:intsed}).

The left-hand panel of Figure~\ref{fig:MS} shows that the stellar masses of our 
WFC3 IR-grism selected galaxies extend over a full three orders of magnitude, with
$7.2 < \log M/M_{\odot} <  10.2$ \citep{gebhardt+15}.  Similarly, the derived SFRs
of our systems span a wide range, with $ 1 \, M_{\odot}$~yr$^{-1} \lesssim {\rm SFR} \lesssim 100 \, 
M_{\odot}$~yr$^{-1}$ \citep{zeimann-1}.  This spread provides leverage for 
our subsequent analysis of the systematic trends in internal reddening.

\subsection{Intrinsic Spectral Energy Distributions}
\label{sec:intsed}

The galaxies selected via our {\sl HST\/} WFC3 IR-grism observations share a common 
characteristic: if photo-ionization from hot stars is responsible for the emission
seen from H$\beta$, [O~III], and [O~II], then the galaxies must have
experienced a period of vigorous star formation within the last 10 Myr
\citep{kennicutt+12}.  Of course, the emission lines do not reveal how long star 
formation has been occurring nor how the rate of star formation has changed with time.  
We can, however, obtain a first order estimate of the star formation timescale
by comparing a galaxy's stellar mass to its current rate of star formation.  
Under the assumption of a constant SFR, the inverse of the 
quantity of mass-specific SFR (sSFR) is simply the system's age.   As shown in 
the right-hand panel of Figure~\ref{fig:MS}, these effective ages range from $\sim 10$ to 
$\sim 500$~Myr, with the sample average around 100 Myr, which is also consistent with 
the ages derived from the stellar mass fits in \S~\ref{sec:masssfr}.  Both stellar population models, 
which assume that stars form stochastically in clusters \citep{dasilva+12}, and 
high-resolution hydrodynamical cosmological simulations \citep{hopkins+14} predict that 
the scatter in the rate of star formation (for SFR $\gtrsim\, 1 \, M_{\odot}$~yr$^{-1}$) , smoothed over bins of $10^7$ years, is typically 
less than a factor of three, and, for the $z \sim 2$ epoch, is more-or-less constant over
$\sim 10^9$ years.  Thus, for the SFRs under consideration (1 to $100  \,M_{\odot}$~yr
$^{-1}$),  the assumption that star formation has been proceeding at a constant rate for
between 10 and 500~Myr is reasonable. 

Of course, alternative star formation histories are certainly viable.  For example, in their study of 302 galaxies in the redshift range $1.5 < z < 2.6$, \citet{reddy+12} found that the best way to reconcile the star formation rates derived from SED modeling with those computed from UV$+$mid-IR flux observations is through an exponentially rising star formation rate.  However, such details have very little effect on our analysis.  For example, if we fit the observed SEDs to models with an exponentially rising star formation rate for a maximally old stellar population (at $z = 2.1$, this corresponds to an age of $\approx$3 Gyr), then the best-fitting e-folding time is $\tau \sim 100$~Myr, and the resulting UV SED (between $1250 < \lambda < 3500$~\AA) is within 1\% of that of a constant SFR model.  In other words, though our choice of star-formation rate history can change the age of the oldest stars in our $z \sim 2$ galaxies, it has essentially no effect on our SED-based analysis of extinction.

For consistency with our calculation of stellar mass, we modeled the galactic SEDs using
the 2007 version of the population synthesis models of \citet{bc03},
with the assumptions of a metal abundance of 
$0.2 \, Z_{\odot}$ \citep{gebhardt+15}, and a \citet{kroupa+01} IMF\null.  Nebular continuum 
emission, which can be an important contributor to the broadband SED of young, low-mass galaxies 
\citep[\eg][]{izotov+11}, was modeled following the prescription of \citet{acquaviva+11} with 
updated templates from \citet[][private communication]{acquaviva+12}.  Nebular line emission was informed by our grism spectroscopy and subtracted from the photometric data as described in \S~\ref{sec:photometry}.

\begin{figure}[t]
\epsscale{0.76}
\plotone{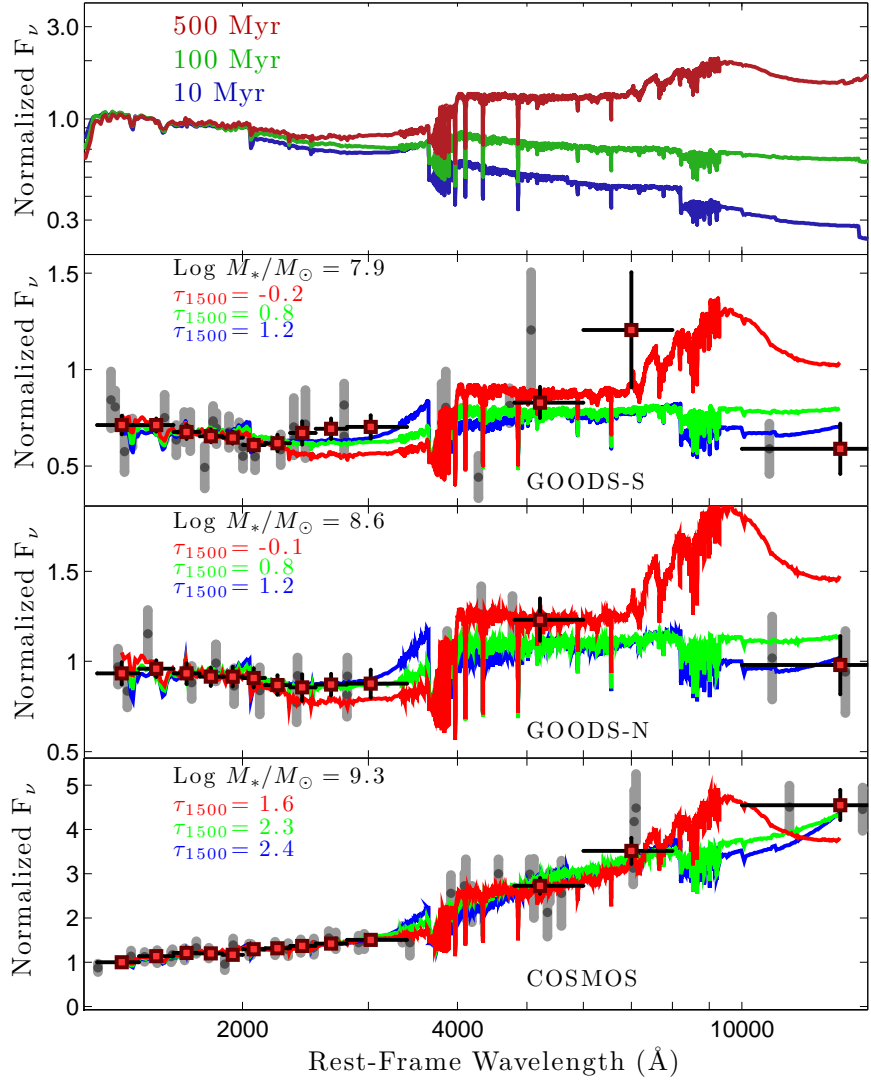}
\caption{The top panel shows 2007 versions of \citet{bc03} SEDs for populations which have been 
forming stars at a constant rate for 10, 100, and 500~Myr.  The data have been arbitrarily 
normalized at 1500~\AA\null.  There is little difference between the populations
in the UV; it is only in the optical and near IR where population age becomes important.
The bottom three panels display three typical rest-frame SEDs from our emission-line 
galaxy sample.  The gray bars are the de-redshifted observed filters, the red squares are the 
``artificial'' filters with error bars illustrating the bin size, 
and the three colored lines correspond to best-fit, dust-attenuated 
model spectral energy distributions (same colors as the top plot).  The attenuation curves 
for the best-fit models are given in Table~\ref{tab:fake filters}, and in each panel the best-fit
${\tau}_{\rm 1500}$ value is reported.  The negative ${\tau}_{\rm 1500}$ values associated with the 500~Myr SED fits for the two lower mass systems illustrate 
that the assumed population is too old, resulting in unphysical attenuation laws.  This may be due to the treatment of asymptotic giant branch stars and is discussed in 
\S~\ref{sec:spssys}.}
\label{fig:stellarsed}
\end{figure}

The top panel of Figure~\ref{fig:stellarsed} 
presents the results of these runs (arbitrarily normalized at 1500~\AA)  for three ages:  10~Myr, 
100~Myr, and 500~Myr.   The most important feature is the similarity of the spectral 
slopes in the UV\null.  As a system ages, the strength of the Balmer break changes, as 
does the ratio of UV to optical light.  However, as quantified by \citet{calzetti01}, there is 
little change in the shape of the stellar continuum between 1250 and 3000~\AA
; on average, the fractional difference between a 10 and 500~Myr is just 8\%.  This provides a 
baseline from which to examine the effects of attenuation in galaxies of different masses and 
star formation rates.     

\section{Uniform Framework for Rest-frame Photometry}
\label{sec:photometry}

Our sample of 239 star-bursting galaxies is distributed  
between $1.90 < z < 2.35$.  Consequently, in the rest frame, the \citet{skelton+14}
photometry of the COSMOS, GOODS-N, and GOODS-S regions sample each SED 
differently.  Our first task was therefore to resample each galaxy's SED onto a standard 
grid in rest-frame wavelength.

To create this information, we began by removing [O~II] $\lambda 3727$,
H$\beta$ and [O~III] $\lambda\lambda 4959,5007$ nebular emission from the \citet{skelton+14} 
photometry using the line flux measurements of \citet{gebhardt+15}.  We also removed H$\alpha$ emission assuming a constant ratio of H$\alpha/$[O~III] $=$ 0.8, which is the expected value for the median metallicity and extinction of our sample, 
and excludef all photometric data with a central wavelength blueward of 1250~\AA, 
where the contribution from Ly$\alpha$ emission is uncertain.  
We then defined 10 artificial rectangular rest-frame filters of intermediate bandwidth, 
spaced to encapsulate an equal number of contributing observed frame filters between 
1250~\AA\ and 3500~\AA\null.   The central wavelengths and widths of these filters are listed in Table~\ref{tab:fake filters}, 
along with those of three additional bandpasses that sample the SED redward of the Balmer 
break and roughly covered the observer's frame H-band, K-band, and the combination of [3.6] and [4.5]$\mu$m filters, respectively.  
We then computed the galaxies' 
stellar flux densities in these artificial bandpasses by interpolating the actual
\citet{skelton+14} photometry onto this new wavelength grid.  Specifically, we calculated
the fractional throughput of a de-redshifted \citet{skelton+14} filter that overlapped 
our artificial rest-frame filter and used that as a weight, along with the inverse variance 
associated with the measurement of flux density.  For a galaxy observed with a set of $S$ filters,  
the weighted flux in the $i^{\rm th}$ artificial filter is then
\begin{equation}
F_{\nu,i} = \frac{\displaystyle \sum_{j \in S} F_{\nu,j} \times w_j / \sigma_{F_{\nu,j}}^2}
{ \displaystyle \sum_{j \in S} w_j /  \sigma_{F_{\nu,j}}^2}
\end{equation}
where $F_{\nu,j}$ is the observed flux density for the $j^{\rm th}$ de-redshifted
observed filter, $\sigma_{F_{\nu,j}}^2$ is the variance of the flux density for the 
observed filter, and $w_j$ is the fraction of the $j^{\rm th}$ de-redshifted observed 
filter that overlaps the $i^{\rm th}$ artificial rest-frame filter.

In general, the rest-frame UV of our {\sl HST\/} grism-selected galaxies is extremely
well sampled by the \citet{skelton+14} photometry, with over two dozen photometric bandpasses 
in COSMOS and GOODS-S
and 12 different filters in GOODS-N\null.  Consequently, by coarsely sampling the UV with
only 10 artificial filters, we are, in effect, smoothing our SEDs as we place each
galaxy on a uniform system.  Note, however, that our resampling algorithm is quite generic, 
and since many of the \citet{skelton+14} bandpasses overlap, we could have chosen to 
resample the observations onto a much denser wavelength grid, i.e.,  to ``drizzle'' the data 
to a higher spectral resolution \citep{fruchter+02}.  However, since high 
frequency features are not expected to be present in the dust attenuation law, we chose to 
improve the signal-to-noise of each data point by keeping our wavelength grid relatively coarse. 
To check this, we repeated our analysis with several different bin sizes ranging from 50~\AA\ to 500~\AA;
in all cases, the results of our analysis were unaffected by the choice of bandpass.

\section{Determining the Dust Attenuation Law}
\label{sec:dustlaw}

In most studies of the SEDs of high-redshift galaxies, one simultaneously fits for a galaxy's
stellar population and internal galactic reddening by making a priori assumptions about
the shape of the dust attenuation curve.  In our case, the mean spectral energy distribution for our vigorously star-forming systems is well enough constrained so that no assumptions
about the behavior of UV dust attenuation are needed, except that its wavelength dependence 
is the same for every galaxy.   In other words, at each point in the wavelength 
grid, the observed flux of a galaxy, $F(\lambda)_{\rm obs}$, is related to its intrinsic
(de-reddened) flux, $F(\lambda)_{\rm int}$, by
\begin{equation}
F(\lambda)_{\rm obs} = F(\lambda)_{\rm int} \, e^{-\tau(\lambda)} = F_0 \, S(\lambda) \,
e^{-\tau_{1500} \left\{ \tau(\lambda)/\tau_{1500} \right\} }
\label{eq:attenuation}
\end{equation}
where $\tau(\lambda)$ is the wavelength dependence of the attenuation normalized
at 1500~\AA, $\tau_{1500}$ is the optical depth of the extinction at 1500~\AA, $S(\lambda)$
is the intrinsic shape of the spectral energy distribution, again normalized at 1500~\AA, and
$F_0$ is the galaxy's total intrinsic flux density at this normalization wavelength.
As illustrated by Figure~\ref{fig:stellarsed}, the UV portion of $S(\lambda)$ should be similar
for every galaxy in our sample.  Thus, equation~(\ref{eq:attenuation}) 
represents a network of 3107 separate equations: (one for every photometric measurement of 
every galaxy) with 491 unknowns (each galaxy's value of $F_0$ and $\tau_{1500}$, 
plus the 13-point measurements of the extinction curve, $\tau(\lambda)$).   (Actually, for a 
few galaxies, some photometric data are missing, so the true total number of equations is 
only 3009.)  The problem is therefore 
highly over-constrained and can be solved via non-linear least squares minimization using
a Levenberg-Marquardt algorithm (such as the MATLAB program {\tt solve}).

For the assumed shape of the intrinsic SED, we adopted our 2007 version of the
\citet{bc03} model for a system that has been forming stars at a constant rate for 100~Myr.
This is the mean age implied by the mass-specific star formation rates shown in 
Figure~\ref{fig:MS}, and is therefore appropriate for the bulk of the galaxies of our
sample.   To examine the possible systematic uncertainty associated with this 
assumption, we also considered two extreme models:  one for a much younger 
system, where star formation began only 10~Myr years ago, and one where steady
star formation has been ongoing for 500~Myr.   These two SEDs should bracket the 
age range of our data and provide an upper limit to whatever systematic
error is associated with our analysis.

\begin{figure}[t]
\epsscale{0.93}
\plotone{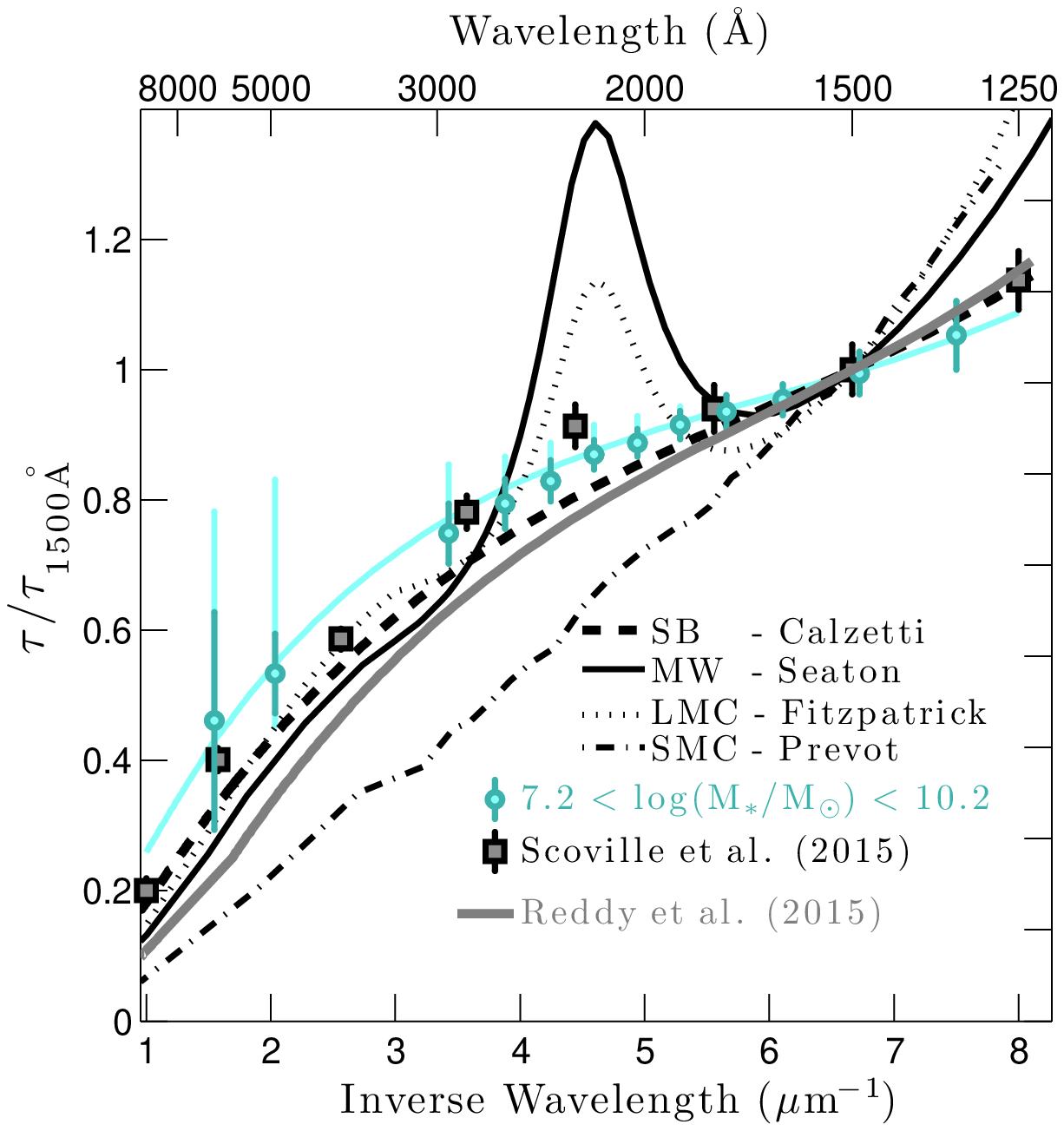}
\caption{Dust attenuation and extinction curves for the Milky Way \citep{seaton+79}, Magellanic Clouds \citep{prevot+84,fitzpatrick+86}, the 
local star-bursting galaxies of \citet{calzetti+00}, the more massive $1.5 < z < 4$ systems 
of \citet{scoville+15} and \citet{reddy+15}, and our sample of rest-frame optical 
emission-line galaxies.  
The dark blue error bars on our data indicate the statistical uncertainties for the dust attenuation 
curve, under the assumption of a 100~Myr constant star-forming stellar population.
The light blue error bars show the extent of the systematic errors if we were to instead 
use our extreme models for 10~Myr and 500~Myr stellar populations.  The blue solid curve indicates
the \citet{calzetti+00} attenuation law, perturbed by our best-fit values of 
$\delta = 0.18 \pm 0.05$ and $E_b = 0.04 \pm 0.27$.}
\label{fig:dustall}
\end{figure}

Figure~\ref{fig:dustall} presents the results of our least squares fit to the 100~Myr
constant SFR model.  From the figure, it is clear that our curve agrees remarkably 
well with that derived by \citet{scoville+15} from a sample of 135 
spectroscopically-confirmed Lyman-break galaxies in the redshift range $2 < z < 4$.   In general,
the masses of the \citet{scoville+15} objects ($M \gtrsim 10^{10} \, M_{\odot}$) are
greater than those of our emission-line selected galaxies, yet the only significant 
difference between the two extinction relations is at 2175~\AA, 
where the higher-mass systems exhibit a small extinction bump.  
Both curves also agree reasonably well with the \citet{calzetti+00} prescription for local starburst 
galaxies \citep[as does the UV portion of the reddening curve derived by][]{reddy+15}, although our 
derived $z \sim 2$ curve has a shallower UV slope. 

Following \citet{kriek+13}, we can quantify the shape of our attenuation curve via a
perturbation on the \citet{calzetti+00} law.  Under this formalism, which was first
explored by \citet{noll+09}, our mean attenuation curve is defined by two parameters,
one reflecting the slope of the reddening relation ($\delta$),  and the other measuring the
amplitude of the 2175~\AA\ extinction bump ($E_b$).  Specifically,
\begin{equation}
\displaystyle \tau(\lambda) = A \times  \Big\{ k(\lambda) + D(\lambda) \Big\}
\left(  \frac{\lambda}{\textrm{1500 \AA}}   \right)^{ \Large \delta}
\label{eq:tau}
\end{equation}
where $k(\lambda)$ represents the wavelength dependence of the 
\citet{calzetti+00} law, $A$ is the curve's overall normalization, and
$D(\lambda)$ is  a Lorentzian-like ``Drude'' profile for the extinction bump
\begin{equation}
D(\lambda) = E_b \, \frac{\left( \lambda \Delta\lambda \right)^2}{\left( \lambda^2 - \lambda_0^2
\right)^2 + \left( \lambda \Delta\lambda \right)^2}
\label{eq:bump}
\end{equation}
with $\lambda_0 = 2175$~\AA\ and $\Delta \lambda = 350$~\AA\  \citep{noll+09}.  
By definition, the wavelength dependence derived by \citet{calzetti+00} for local
starburst galaxies has $E_b = \delta = 0$.  For our best-fit curve in Figure~\ref{fig:dustall}, 
$\delta = 0.18 \pm 0.05$, 
and there is no hint of a 2175~\AA\ bump 
($E_b = 0.04 \pm 0.27$).  For reference, the screen-model extinction law produced by 
Milky Way dust has $E_b \sim 4.3$.

\section{The $\MakeLowercase{z} \sim 2$ Attenuation Law as a Function of Stellar Mass}
\label{sec:dustlawmass}

By fitting the individual SEDs of 751 galaxies between $0.9 < z < 2.2$, \citet{buat+12} found
that the dispersion in the best-fit values of $\delta$ and $E_b$ was much larger than could
be explained by photometric errors and outliers.  Moreover, subsets of their data  
occupied different regions of the $\delta$-$E_b$ phase space, suggesting that galaxies with 
different physical parameters possess different attenuation curves.  This result was 
confirmed by \citet{kriek+13}, who reported that the strength of a galaxy's 2175~\AA\ bump and 
the slope of its UV reddening law are both inversely correlated with the equivalent width of its
H$\alpha$ emission line.

As with most samples of low- and high-redshift galaxies, the physical properties of our
IR-grism selected objects (i.e., stellar mass, SFR, metallicity, emission-line equivalent widths,
size, total extinction, etc.)~are strongly correlated (or inversely correlated) with each other 
\citep{gebhardt+15, hagen+15}.  Consequently, by investigating the systematics of dust 
attenuation against any one of these variables, we are, in fact, examining its behavior
against an entire suite of parameters.  Since stellar mass has been found to be the primary 
predictor of many galactic properties, both at high- and low-redshift  \citep[i.e.,][]{tremonti+04,
garn-best, peng+10}, we adopt it as the independent parameter in our study.  

\begin{figure}[t]
\epsscale{0.93}
\plotone{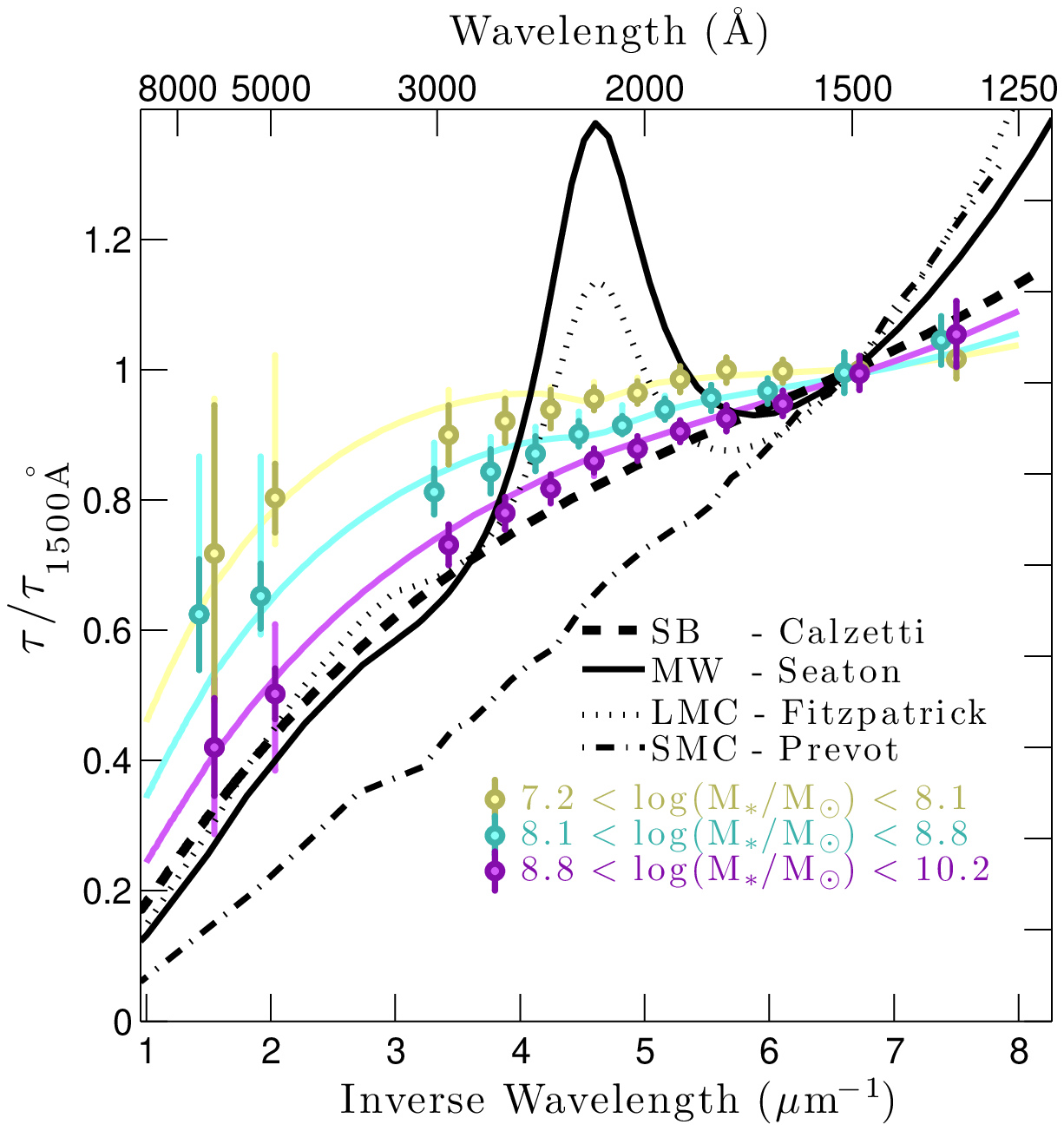}
\caption{Dust attenuation curves for our sample of {\sl HST\/} WFC3 IR-grism selected
galaxies split into three stellar mass bins, as well as the curves for the Milky Way \citep{seaton+79}, Magellanic Clouds \citep{prevot+84,fitzpatrick+86}, and for the local star-bursting galaxies of \citet{calzetti+00}.  The dark color error bars indicate the statistical 
uncertainties for the dust attenuation curve, under the assumption of a 100~Myr constant 
star-forming stellar population; the light color error bars show the extent of the systematic 
errors if we were to instead use our extreme models for 10~Myr and 500~Myr stellar 
populations.  The colored curves display the \citet{calzetti+00}  attenuation law perturbed
by our best-fit values for $\delta$ and $E_b$.  The three mass bins clearly have different 
attenuation laws, with a shallower curve associated with least massive systems.}
\label{fig:dustbin}
\end{figure}

To investigate the dependence of the dust attenuation law on stellar mass, we sub-divided our sample of 239 galaxies into three bins spaced roughly equally in log mass.   
We then measured
each group's reddening law and perturbation parameters ($E_b$ and $\delta$)
in exactly the same fashion as in \S\ref{sec:dustlaw}.   The results are listed in 
Table~\ref{tab:massbins} and displayed in Figure~\ref{fig:dustbin}.   The data clearly
indicate a systematic shift in the dust attenuation curves for different
mass galaxies.   A gradient is apparent in the slopes of the three attenuation curves, 
with the higher mass objects having a steeper curve in the UV.  None of the stellar mass bins show evidence of a 2175~\AA\ bump.   
 
\begin{figure}[t]
\epsscale{1.0}
\plotone{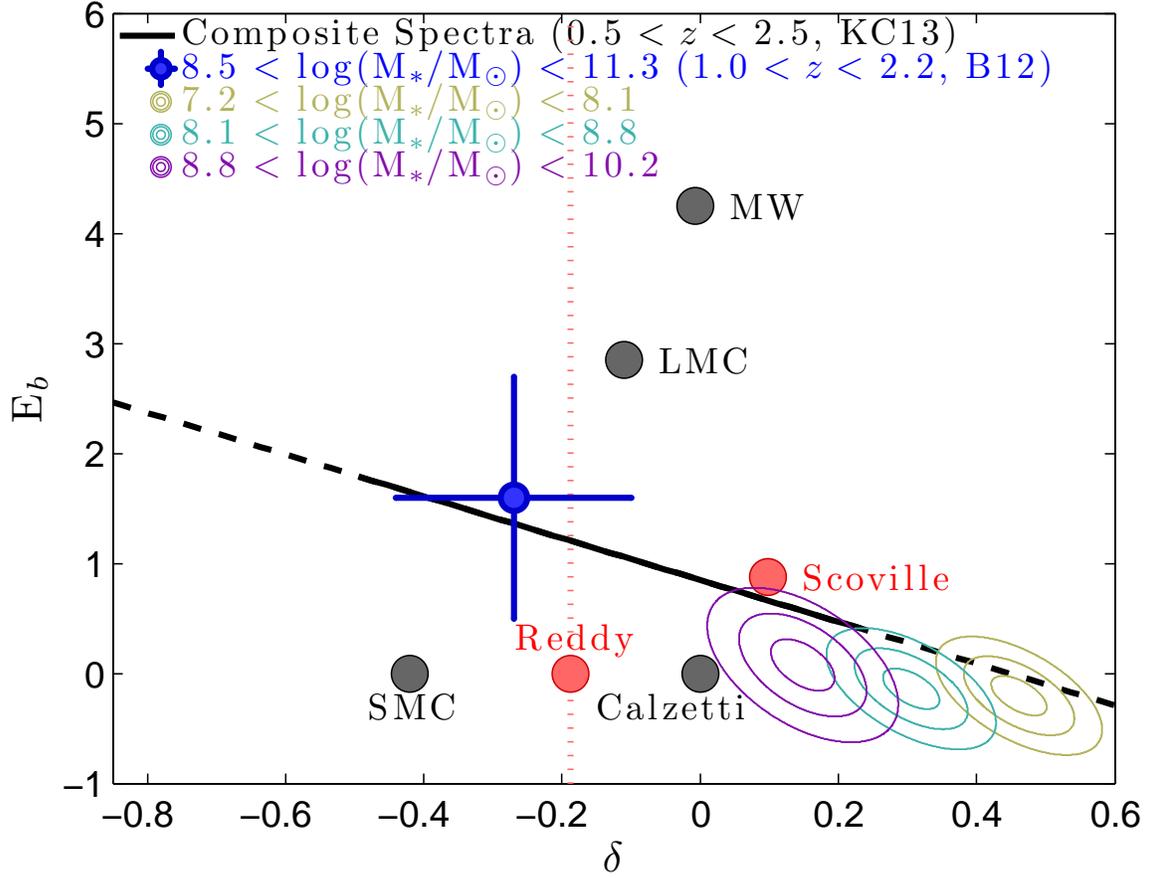}
\caption{Measurements of the dust attenuation law parameters as a function of galaxy stellar
mass for $z \sim 2$ systems.  The variable $\delta$ reflects the UV slope of the
reddening curve; $E_b$ describes the amplitude of the 2175~\AA\ bump.  The
colored ellipses represent $1$, $2$, and $3 \, \sigma$ uncertainties for the three mass bins.  By
definition, a \citet{calzetti+00} law has $E_b = \delta = 0$.  The trend in our
measurements and those of \citet{buat+12} agree qualitatively with that found by
\citet{kriek+13}.   In all cases, the \citet{calzetti+00} curve is ruled out with greater than
$3 \, \sigma$ confidence.  Also shown are the results from \citet{scoville+15} and \citet{reddy+15}, which include galaxies that are more massive, on average, than those studied in this work. By construction, \citet{reddy+15} do not fit for a bump in their dust attenuation curve, so we use a vertical dotted line to illustrate the uncertainty.}
\label{fig:dustcompare}
\end{figure}

Figure~\ref{fig:dustcompare} quantifies the systematic behavior of the attenuation 
curve by plotting the best-fit values of $E_b$ and $\delta$ (and their uncertainties)
against stellar mass.  Also shown are the results from \citet{buat+12}, \citet{scoville+15}, and \citet{reddy+15}, along with the
Milky Way and Magellanic Cloud laws.  The figure demonstrates that the behavior of 
these perturbation parameters are qualitatively consistent with the trends found from the 
composite spectral analysis of \citet{kriek+13}.   Moreover, although the \citet{calzetti+00} law
is superficially similar to our attenuation curve, in every case its far-UV slope is ruled out 
with greater than $3 \, \sigma$ confidence.  This result is true even for the highest mass objects, 
where the reddening law lies closest to that seen in local starburst systems.

We can generalize these results by fitting a line through the data and solving for the
dependence of $\delta$ and $E_b$ on stellar mass.  Our best-fit regressions yield 
\begin{equation}
\displaystyle \delta = (0.19 \pm 0.04) - (0.23 \pm .04)\ [\log \left( M_*/M_{\odot} \right) - 9],
\label{eq:deltamass}
\end{equation}
\begin{equation}
\displaystyle E_b = (0.00 \pm 0.12) + (0.20 \pm 0.15)\ [\log \left( M_*/M_{\odot} \right) - 9].
\label{eq:ebmass}
\end{equation}
Of course, these relations are somewhat sensitive to the systematics of our 
stellar mass measurements.  
For example, our estimates of $M_*$ were made 
by assuming a \citet{kroupa+01} IMF with a 2007 version of a \citet{bc03} constant 
star-formation rate model with $Z = 0.2 Z_{\odot}$.  
Mass estimates based on a \citet{salpeter55} 
initial mass function may be systematically larger by $\sim 0.3$~dex, while solar 
metallicity populations may be larger by $\sim 0.1$.  \citep[See][for a full discussion 
of the systematics of SED fitting.]{conroy13}  However, these shifts will 
only result in a re-scaling of the stellar mass zeropoint in equations~(\ref{eq:deltamass}) 
and (\ref{eq:ebmass}).  Unless the IMF of a galaxy changes with its stellar mass, the
slope of the relation will be unaffected.

\subsection{Intrinsic Stellar Population Systematics}
\label{sec:spssys}

A larger systematic concern is that associated with the assumed intrinsic stellar population, which is set by both the stellar population synthesis (SPS) model 
and the chosen age of the system.  In \S~\ref{sec:intsed}, we showed that our galaxies are vigorously star-forming and have ages between 10 and 500~Myr.  Over 
this age range and under the assumption of a constant star formation history, the largest changes in the assumed intrinsic SED occur in the optical and near-IR, 
while the UV remains relatively unaffected.  With regards to the choice of SPS models, two important factors are the inclusion of nebular continuum 
emission\footnote[5]Most SPS models do not include nebular continuum/line emission;  instead, the application is usually an add-on recipe \citep[\eg]
[]{acquaviva+12}. and the prescription for asymptotic giant branch (AGB) stellar evolution.  While the latter parameter principally affects predictions for the 
flux in the red and near-IR, the former may contribute significantly at both optical and UV wavelengths.  To quantify these effects, we compared the SEDs of 
several of the most common SPS codes, fixing the stellar metallicity to $Z = 0.2\, Z_{\odot}$, and examining the results for constant star-formation rate models 
with ages of 10, 100, and 500~Myr.  Included in this comparison were the 2003 and 2007 versions of BC03 and CB07\citep[][]{bc03}, version 7.0.1 of the 
STARBURST99  \citep[SB99;][]{SB99, vazquez+05, SB99-2, SB99-3}, and version 2.5 of the Flexible Stellar Population Synthesis code 
\citep[FSPS;][]{fsps-1,fsps-2}. 

\begin{figure}[t]
\epsscale{0.76}
\plotone{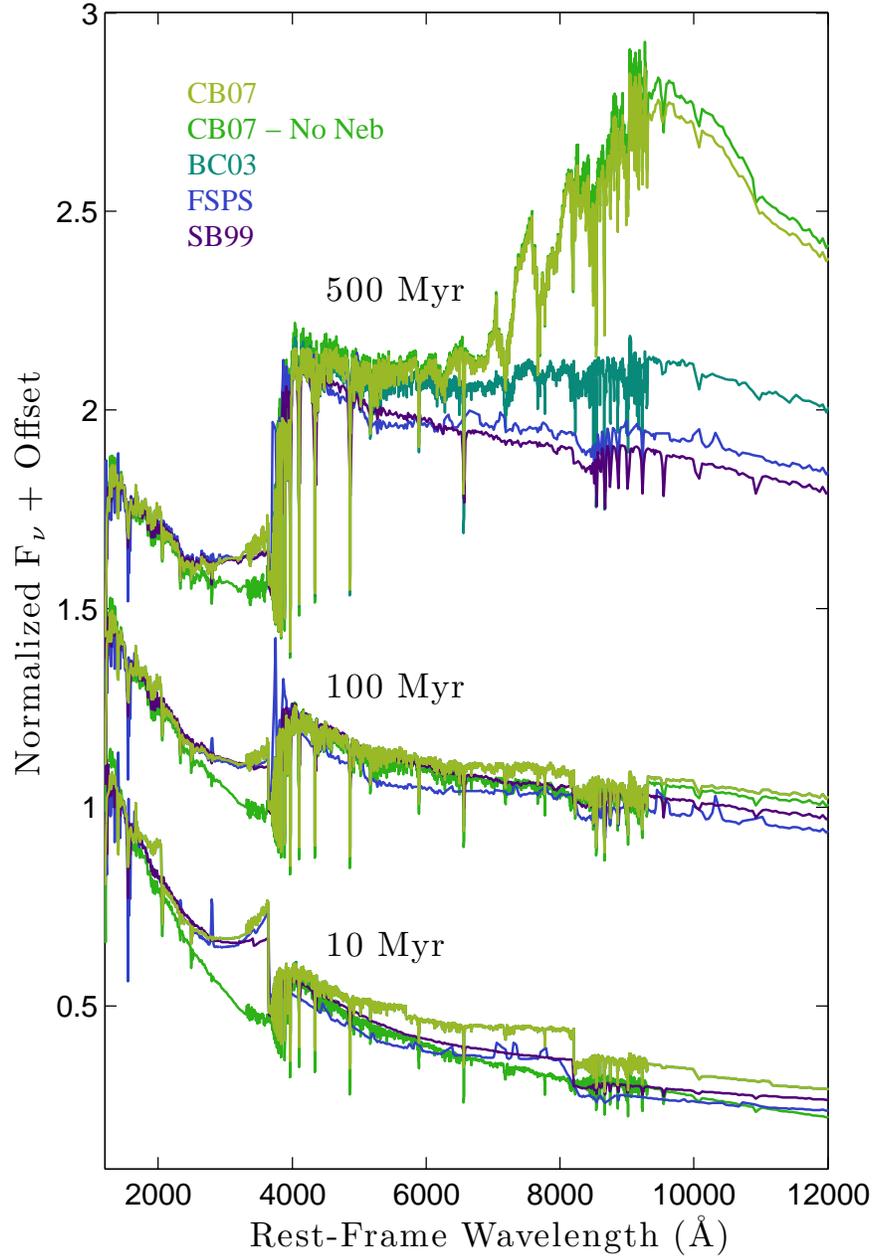}
\caption{A comparison of the spectral energy distributions produced by several popular population synthesis codes.  The data have been arbitrarily normalized at 1500~\AA\null.  We show the SEDs create by the 2007 version of \citet{bc03} with and without nebular continuum emission, the original \citet{bc03} model, version 7.0.1 of STARBURST99, and FSPS.  All of the SPS codes were set to the same stellar metallicity, $Z = 0.2 Z_{\odot}$ and used constant star formation histories with ages of 10, 100, and 500~Myr.  There is little difference between the populations in the UV with the exception of the model without nebular continuum emission; it is only in the optical and near IR where population age and the codes' differing assumptions become important.
}
\label{fig:spsall}
\end{figure}

Figure~\ref{fig:spsall} shows the results of this test.  In the near-UV, the slope of the SED is altered due to nebular continuum emission, which is included or 
added to all of the SPS codes, except for one run with CB07\null.  Nebular continuum emission is clearly most important in the youngest populations, but its 
effect on the near-UV slope is present in systems of all ages.  In the optical and near-IR, the importance of AGB stars increases with age, and the various 
treatments of this component are responsible for the dramatic differences seen at 500~Myr.  Near $1~\mu$m, the differences between the CB07 and SB99 500~Myr 
models are just as large as those between the SB99 SEDs of 10 and 500~Myr.

Despite these uncertainties, even in the most extreme case, where we use 500~Myr models for the most massive galaxies, and allow the least massive systems to be 
10~Myr old, the UV attenuation curves differ with more than $\sim 4\, \sigma$ confidence.  The same results would be obtained if we were to assume solar 
metallicity for the highest mass galaxies and use a $0.2 Z_{\odot}$ model for the lowest mass systems.  The effective attenuation law for a $z \sim 2$ 
emission-line galaxy is clearly a function of its stellar mass, with the higher-mass objects having steeper UV slopes.  Only if the star formation history and dust attenuation law are varied in unison would the strength of the correlation diminish. This possibility will be investigated in more detail in \citet{brooks+16}.

\section{Discussion}
\label{sec:Discussion}

In the sections above, we presented our analysis of galactic attenuation versus stellar 
mass, and gave the variations of $E_b$ and $\delta$ against this parameter.   Comparisons 
to other physical parameters yield similar results.  When we repeat the experiment using
SFR, sSFR, and half-light radius, we see the same trends:  the shape of the dust
attenuation curve goes from a nearly a \citet{calzetti+00} relation for the lowest sSFR, 
highest SFR, and largest galaxies to a shallower curve in the UV for the highest sSFRs, 
lowest SFRs, and smallest systems.  From the data, it is unclear which of these physical 
parameters is most correlated with the trends in the attenuation curve.  Nonetheless, the 
existence of the correlations is quite compelling.

\begin{figure}[t]
\epsscale{0.43}
\plotone{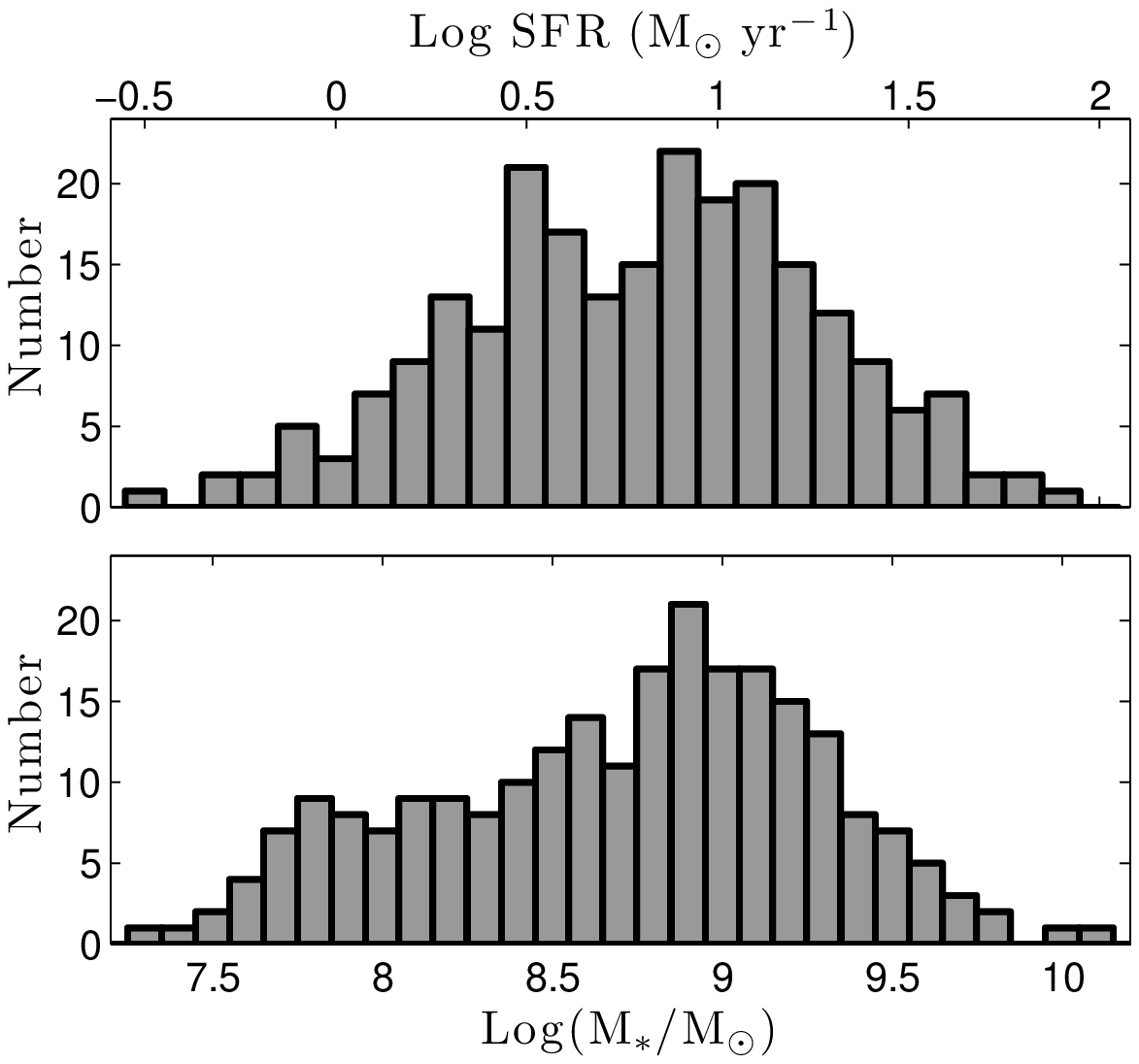}
\epsscale{0.51}
\plotone{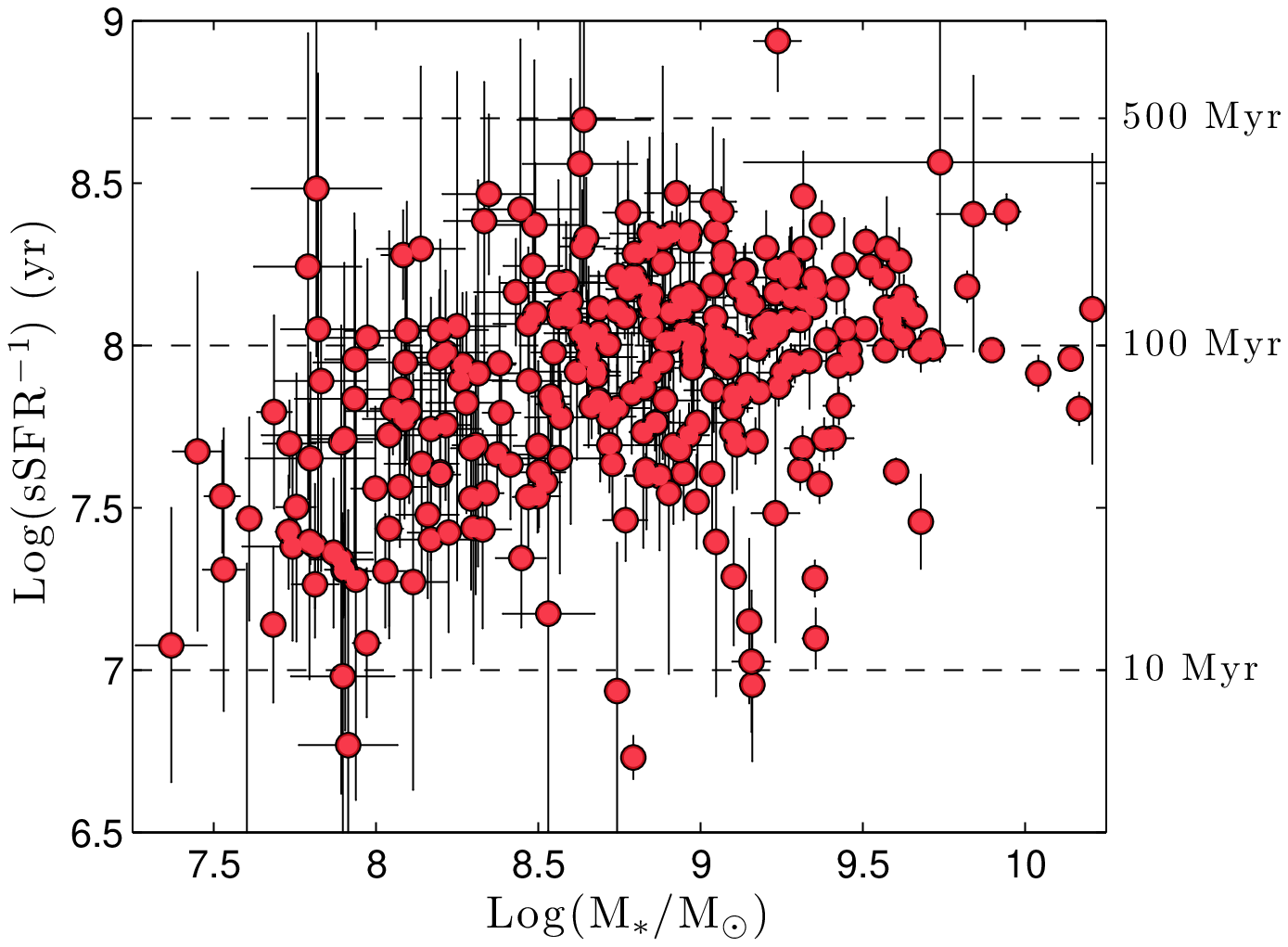}
\caption{{\it Left:} The stellar mass and dust-corrected UV-based SFR distributions of the 239 galaxies in
our {\sl HST\/} WFC3 IR-grism selected sample.  The data span
three orders of magnitude in stellar mass, and two orders of magnitude in SFR, and include
objects that are generally smaller and fainter than galaxies selected via their continuum
colors.  {\it Right:} The mass-specific star formation rates of 
the galaxies as a function of their stellar mass.   The error bars reflect only the
statistical uncertainties associated with our measurements; the systematic offsets 
associated with parameters such as the assumed IMF and SFR calibration are not shown.
For galaxies with a constant star formation rate, the sSFR is simply the inverse of the
age, which is indicated via the right-hand labels of the figure.   The data indicate
that most of our galaxies have likely been forming stars for at least $\sim 100$~Myr. 
The observed correlation betwen sSFR and stellar mass may be affected by the SFR 
limit of the grism survey.  This selection effect is described in more detail in 
\citet{zeimann-1,zeimann-2}.}
\label{fig:MS}
\end{figure}

The mass dependency of our empirically-derived attenuation curves is consistent with the 
age-dependent star-to-dust geometries modeled by \citet{charlot+00}.  As the 
right-hand panel of Figure~\ref{fig:MS} illustrates, lower mass systems in our sample
have systematically younger ages as defined by the stellar mass to SFR ratio.  Therefore, these
lower mass galaxies should have more of their UV light enshrouded in an optically thick dust 
cloud.   This situation should produce a shallower attenuation curve \citep{granato+00}, which is
exactly what is observed.

On the other hand, our low-mass systems also have low metallicities 
\citep{gebhardt+15} and possibly more intense interstellar radiation 
fields (ISRF) due to their higher sSFRs.  The dust carrier of the 2175~\AA\ bump 
may be easily destroyed in such an environment, or perhaps the lack of metals changes
the dust composition, which then affects the wavelength dependence of opacity.
In the SINGS \citep{kennicutt+03} survey, \citet{draine+07} found that the mid-IR 
emission from polycyclic aromatic hydrocarbons drops dramatically for galactic 
metallicities below $12+{\rm log(O/H)} < 8.1$.  Since the origin of the 2175~\AA\ bump 
has been suggested to be carbonaceous in nature, perhaps the two parameters are linked 
\citep{draine11}.  We do not offer a physical model for the dust composition as a 
function of metallicity or ISRF;  we only  state that we cannot exclude it as the 
explanation for our observed correlation between the steepness of the dust attenuation 
curve and stellar mass.

Disentangling the effects of ISM geometry from dust composition is an extremely
difficult undertaking \citep{penner+15}.  Ideally, one would require high-spatial resolution imaging to observe
both the stars and individual H~II regions and directly determine the star-dust geometry.
Unfortunately, at $z \sim 2$, this would require $\sim 10$~pc, or $\sim 0 \farcs 001$ 
resolution, which is well beyond current technology.  At the same time, one would also
like to probe the chemistry of both the gas and the dust via their emission
features.  Again, this task is exceedingly difficult, as it requires deep optical, near-IR,
and mid-IR spectrophotometry.  Although the optical and near-IR observations are 
theoretically possible using instruments like MOSFIRE \citep{mosfire}, it is time-prohibitive
for the lowest mass objects in our sample.

Nevertheless, the applicability of our results is far-reaching.  Emission-line selected galaxies,
such as those analyzed here, are poised to quickly become the dominant population of 
known $z \gtrsim 1$ objects.  Extremely large, wide-field cosmological surveys such as 
HETDEX \citep[Hobby Eberly Telescope Dark Energy Experiment;][]{HETDEX} and Euclid \citep{EUCLID} 
will construct samples of millions of 
emission-line galaxies, via the detection of Ly$\alpha$, H$\alpha$, [O~II], and
[O~III]\null.  In order to exploit these data for galaxy evolution studies, 
one will need to understand the relation between objects' intrinsic and observed SEDs.
This work is a step in that direction.

\acknowledgments
We would like to thank V. Acquaviva for her help and support of {\tt GalMC}.  This work was 
supported via NSF through grant AST 09-26641\null.  The  Institute for Gravitation and the Cosmos 
is supported by the Eberly College of  Science and the Office of the Senior Vice President for 
Research at the  Pennsylvania State University.  

{\it Facilities:} \facility{HST (WFC3)}

\begin{deluxetable}{ccccc}
\centering
\tabletypesize{\normalsize}
\tablecaption{Best-Fit Attenuation Curve}
\tablewidth{0pt}
\tablehead{
 \colhead{$\lambda_{\rm c}$\tablenotemark{1}} &  \colhead{$\Delta(\lambda)$}
 &\colhead{100 Myr SED} & \colhead{10 Myr SED} & \colhead{500 Myr SED}\\
 \colhead{(\AA)} &  \colhead{(\AA)} & \colhead{$\tau(\lambda) / {\tau}_{\rm 1500}$} &
 \colhead{$\tau(\lambda) / {\tau}_{\rm 1500}$} &
 \colhead{$\tau(\lambda) / {\tau}_{\rm 1500}$}
 }
 \startdata
 1341 & 182   & $1.053 \pm 0.054$ & $1.044 \pm 0.042$ & $1.032 \pm 0.084$\\
 1516 & 168   & $0.995 \pm 0.033$ & $0.996 \pm 0.027$ & $0.997 \pm 0.039$\\
 1671 & 141   & $0.954 \pm 0.024$ & $0.949 \pm 0.019$ & $0.968 \pm 0.039$\\
 1807 & 132   & $0.935 \pm 0.026$ & $0.935 \pm 0.021$ & $0.957 \pm 0.026$\\
 1938 & 129   & $0.915 \pm 0.022$ & $0.923 \pm 0.019$ & $0.946 \pm 0.026$\\
 2076 & 147   & $0.887 \pm 0.021$ & $0.873 \pm 0.018$ & $0.931 \pm 0.026$\\
 2239 & 181   & $0.869 \pm 0.023$ & $0.856 \pm 0.019$ & $0.918 \pm 0.032$\\
 2431 & 201   & $0.828 \pm 0.032$ & $0.819 \pm 0.026$ & $0.891 \pm 0.040$\\
 2664 & 267   & $0.793 \pm 0.039$ & $0.786 \pm 0.033$ & $0.872 \pm 0.049$\\
 3099 & 602   & $0.747 \pm 0.046$ & $0.764 \pm 0.043$ & $0.860 \pm 0.056$\\
 5400 & 1200  & $0.530 \pm 0.062$ & $0.450 \pm 0.053$ & $0.833 \pm 0.086$\\
 7000 & 2000  & $0.457 \pm 0.168$ & $0.369 \pm 0.145$ & $0.789 \pm 0.177$\\
 14000 & 8000 & $0.315 \pm 0.116$ & $0.127 \pm 0.118$ & $0.781 \pm 0.222$\\
 \hline
\multicolumn{2}{l}{Reduced $\chi^2$} & 1.26 & 1.18 & 4.46\\
\multicolumn{2}{l}{Reduced $\chi^2$ ($\lambda < 3500$~\AA)} & 1.03 & 0.96 & 2.16
\enddata
\tablenotetext{1}{Central Wavelength of the artificial filters.}
\label{tab:fake filters}
\end{deluxetable}

\begin{deluxetable}{lcccccr}
\centering
\tabletypesize{\footnotesize}
\tablecaption{Best-Fit Attenuation Curves for Different Stellar Mass Bins}
\tablewidth{0pt}
\tablehead{
\colhead{Stellar Mass Bin} & \colhead{$\langle \log M_*/M_{\odot} \rangle$}
& \colhead{\# of gals.} & \colhead{\# of eqns.} & \colhead{Reduced $\chi^2$}
&\colhead{$\delta$} & \colhead{$E_b$}
 }
 \startdata
$7.2 < \log M_* < 8.1$  &7.83 & 39  & 457   & 0.85 & $0.459\pm0.040$ & $-0.21\pm0.17$\\
$8.1 < \log M_* < 8.8$  &8.52 & 81  & 1015  & 0.88 & $0.307\pm0.039$ & $-0.15\pm0.19$\\
$8.8 < \log M_* < 10.2$ &9.22 & 119 & 1537  & 1.39 & $0.141\pm0.049$ & $0.09\pm0.23$
\enddata
\label{tab:massbins}
\end{deluxetable}
\clearpage

\clearpage
\end{document}